\documentclass[a4paper,fleqn,usenatbib]{mnras}
\usepackage{newtxtext,newtxmath,hyperref}
\usepackage[T1]{fontenc}
\usepackage{ae,aecompl}
\usepackage{graphicx}	
\usepackage{amsmath}	
\usepackage{amssymb}	
\usepackage{url}
\usepackage{bm}

\if{
documentclass[a4paper,fleqn,usenatbib]{mnras}
\bibliographystyle{apj}
\usepackage{ae,aecompl}
\usepackage{graphicx}

\usepackage{amsmath}	
\usepackage{amssymb,color}	
}\fi

\newcommand{\Msun}{\ensuremath{\mathrm{M}_\odot}}

\if{
\slugcomment{draft v1}
\shorttitle{Polarization of early-time Macronova}
\shortauthors{T. Matsumoto}

\begin{document}
\title{Polarization of the first hour macronova}

\author{Tatsuya Matsumoto\altaffilmark{1,2,3}}
\email{tatsuya.matsumoto@mail.huji.ac.il}

\altaffiltext{1}{Racah Institute of Physics, Hebrew University, Jerusalem, 91904, Israel}
\altaffiltext{2}{Department of Physics, Graduate School of Science, Kyoto University, Kyoto 606-8502, Japan}
\altaffiltext{3}{JSPS Research Fellow}
}\fi

\title[Polarization of the first-hour macronovae]{Polarization of the first-hour macronovae}

\author[T. Matsumoto]{Tatsuya Matsumoto$^{1,2,3}$\thanks{E-mail: tatsuya.matsumoto@mail.huji.ac.il}
\\
$^{1}$Racah Institute of Physics, Hebrew University, Jerusalem, 91904, Israel\\
$^{2}$Department of Physics, Graduate School of Science, Kyoto University, Kyoto 606-8502, Japan\\
$^{3}$JSPS Research Fellow
}
\pubyear{2018}

\begin{document}
\label{firstpage}
\pagerange{\pageref{firstpage}--\pageref{lastpage}}
\maketitle

\begin{abstract}
Macronovae (or kilonovae) are the optical and NIR counterparts of binary neutron star mergers.
While the macronova in GW170817 was detected about 10 hours after the GW detection, future observations can possibly detect them within the first hour after the merger.
Early-time macronovae are potentially powered by some mechanisms such as the beta-decay heating of the surviving free neutrons.
In this paper, we propose that polarimetric observation can be a useful tool to study the early macronova emissions.
If free nucleons remain in the outermost layer of the ejecta, the electron scattering produces a larger polarization than that by the \textit{r}-process element-rich ejecta.
The degree of polarization can show a large value of $\sim1-3\,\%$ for the first $0.3-1\,\rm{hr}$ for free nucleon mass of $10^{-5}-10^{-4}\,\Msun$.
Quick polarimetric observations enable us to study not only the aspherical morphology of the ejecta but also the amount of the free nucleons in the ejecta, which is helpful to discriminate the emission mechanisms of the early macronovae.
\end{abstract}


\section{INTRODUCTION}\label{intro}
On 17th August 2017, the LIGO \& VIRGO collaboration detected the gravitational waves (GWs) from a binary neutron star (NS) merger for the first time \citep[GW170817,][]{Abbott+2017c}.
After the GW detection, many observation groups carried out the electromagnetic-counterpart searches all over the world \citep{Abbott+2017_gw170817_multi}, and they detected a gamma-ray flash \citep{Abbott+2017e,Goldstein+2017,Savchenko+2017}, a UV, optical, and NIR emissions \citep{Arcavi+2017,Chornock+2017,Coutler+2017,Cowperthwaite+2017,Diaz+2017,Drout+2017,Evans+2017,Kasliwal+2017,Kilpatrick+2017,McCully+2017,Nicholl+2017,Pian+2017,Shappee+2017,Smartt+2017,Soares-Santos+2017,Tanvir+2017,Tominaga+2018,Utsumi+2017,Valenti+2017}, and X-ray and radio afterglows \citep{Haggard+2017,Margutti+2017,Troja+2017,Alexander+2017,Hallinan+2017,Kim+2017}.

Theoretically, optical and NIR counterparts of binary NS mergers have been predicted to be produced by the radioactive decay heating of heavy elements synthesized in the ejecta, and they are named as macronovae \citep{Li&Paczynski1998,Kulkarni2005} or kilonovae \citep{Metzger+2010}.\footnote{In addition to the \textit{r}-process element decay heating, central-engine activities can also power emissions like macronovae \citep{Kisaka+2015,Kisaka+2016,Matsumoto+2018b}. Therefore, we call a transient associated with binary NS mergers as macronovae whatever the energy source is, like ``supernovae''.}
In neutron-rich ejecta, the neutron-capture reaction advances rapidly and synthesizes the \textit{r}-process elements.
These products are unstable and decay to heat up the ejecta.
Detailed emission models have been studied based on many numerical simulations on hydrodynamics of the ejecta \citep{Hotokezaka+2013,Dessart+2009,Perego+2014,Sekiguchi+2015,Sekiguchi+2016,Fujibayashi+2017,Fujibayashi+2018,Fernandez&Metzger2013,Metzger&Fernandez2014,Fernandez+2015,Siegel&Metzger2017,Kiuchi+2014,Giacomazzo+2015,Ciolfi+2017}, the nucleosynthesis \citep{Wanajo&Janka2012,Wanajo+2014,Just+2015,Lippuner+2017}, and the radiative transfer \citep{Kasen+2013,Tanaka&Hotokezaka2013,Tanaka+2018}.
The emission models are also applied to GW170817 \citep{Kasen+2017,Tanaka+2017,Shibata+2017b,Kawaguchi+2018}.

In the up-coming GW observations, the counterpart searches will be started in a shorter timescale than an hour (for GW170817, the detection was announced in the GCN circular $\sim40\,\rm{min}$ after the GW observation), while the macronova in GW170817 was discovered $\sim10\,\rm{hrs}$ after the GW detection \citep{Coutler+2017}.
The first-hour macronovae can be helpful to constrain the emission models of macronovae \citep{Arcavi2018}.
Furthermore, since the early emission is potentially powered by other mechanisms rather than the \textit{r}-process heating, the observations of them may give us the information on the \textit{r}-process nucleosynthesis or activities of the merger remnants.
For example, if the nucleosynthesis does not advance in the fast expanding outer layer of the ejecta and leaves free neutrons, their beta-decay produces a bright emission \citep[so-called, a neutron precursor,][]{Metzger+2015}.
Moreover, if a relativistic jet is launched by the merger remnant, it forms a cocoon during the propagation in the ejecta \citep{Nakar&Piran2017}.
When the cocoon breaks out of the ejecta, its cooling emission also contributes to the early emission \citep{Gottlieb+2018,Wang&Huang2018}.

In this paper, we propose that early-time polarimetric observations can be a powerful tool to study the first-hour macronovae.
As we explain in section \ref{fast velocity tail}, if free nucleons survive in the outermost layer of the ejecta, the electron scattering dominates the opacity, which produces a larger degree of polarization than that produced by the \textit{r}-process element-rich ejecta.
We study the temporal evolution of the degree of polarization by constructing a simple model and show that the polarimetric observations enable us to study not only the morphology of the ejecta, but also to estimate the mass of the free nucleons that survived the \textit{r}-process nucleosynthesis.
The early polarimetric observations can be a probe to study the condition on the \textit{r}-process nucleosynthesis or even discriminate the emission mechanisms of the early macronvae.

We organize this paper as follows.
In the next section, we describe our model.
First, we review the basic properties of the fast expanding ejecta where nucleons survive (section \ref{fast velocity tail}).
Next, we explain how to calculate the degree of polarization produced in the early macronovae (section \ref{polarization}).
Then, we construct a light curve model taking the \textit{r}-process element and neutron decay heating into account (section \ref{light curve}).
The resulting light curves and temporal behaviors of the polarization are shown in section \ref{results}.
We finally summarize this work and discuss implications to observations in section \ref{summary and discussion}.

\section{Model}\label{model}
We construct a simple model to study the first-hour macronovae and temporal evolution of the degree of polarization.
To compare the results with the observed macronova in GW170817, we also consider the late-time ($>\,\rm{hr}$) macronova component.
The macronova in GW170817 showed a transition from optical ($\sim\rm{\,days}$) to NIR emissions ($\sim\,\rm{weeks}$), which are called blue and red macronovae, respectively.
In this work, we include only the blue macronova component, and do not consider the red macronova emission.

\subsection{Fast velocity tail}\label{fast velocity tail}
We discuss properties of a fast expanding outermost layer of the ejecta.
This component is called as the fast velocity tail, which is the fastest ejecta produced just after the coalescence of a binary NS.
Originally, this component was recognized in numerical simulations as ejecta with velocity $\gtrsim0.5\,c$ and mass $\lesssim10^{-4}\,\Msun$ \citep{Hotokezaka+2013,Bauswein+2013,Metzger+2015,Hotokezaka+2018b}, where $c$ is the speed of light.
One of the possible mechanisms producing the fast tail is the shock breakout from a contact surface of the merged NS \citep{Kyutoku+2014,Ishii+2018}.
When a shock wave emerges from the surface of the NS, it accelerates a small amount of the stellar matter even to a relativistic velocity.
The ejecta mass and velocity are smaller and larger than those of the blue and red macronvae of $\sim10^{-3}-10^{-2}\,\Msun$ and $\sim0.1-0.3\,c$.

A fast velocity tail possibly shows an emission in an hour-timescale after the merger.
The fast tail is ejected first and located at the head of the ejecta.
Due to the small ejecta mass $\lesssim10^{-4}\,\Msun$, the fast tail has a short characteristic emission timescale, so-called the diffusion timescale \citep{Arnett1980}, of 
\begin{eqnarray}
t_{\rm{diff}}&\sim&\sqrt{\frac{\kappa{M}}{4\pi{vc}}},
   \label{diffusion time}\\
&\sim&1{\,\rm{hr}\,}\biggl(\frac{\kappa}{1{\,\rm{cm^2\,g^{-1}}}}\biggl)^{1/2}\biggl(\frac{M}{10^{-4}\,\Msun}\biggl)^{1/2}\biggl(\frac{v}{0.5\,c}\biggl)^{-1/2},
\end{eqnarray}
where $\kappa$, $M$, and $v$ are the ejecta opacity, mass, and velocity, respectively.

In the fast velocity tail, the \textit{r}-process nucleosynthesis may not occur efficiently.
In most of the ejecta from a binary NS merger, the \textit{r}-process nucleosynthesis advances and produces heavy elements.
On the other hand, since the fast tail has a smaller expansion timescale than that of the late-time macronova ejecta, the neutron-capture reaction may not proceed efficiently and leaves free nucleons.
\cite{Metzger+2015} analyzed the results of numerical simulations by \cite{Bauswein+2013,Just+2015} and found that if the expansion timescale is smaller than $\sim5\,\rm{ms}$, free neutrons may remain in the fast ejecta.
They also discussed that the beta-decay of the free neutrons powers a precursor emission to macronovae although the final abundance of the neutrons is uncertain.
\cite{Ishii+2018} also carried out numerical simulations of shock breakout from a NS and confirmed that a fraction of neutrons remains in the outermost layer.

When ejecta are composed of nucleons rather than the \textit{r}-process elements, the electron scattering becomes a main opacity source.
Furthermore, if the photosphere has a deviation from the spherical symmetry, the emission shows a continuum linear polarization as actually observed in supernovae (SNe) \citep[for a review]{Wang&Wheeler2008}.
In contrast to the electron scattering, the bound-bound transition dominates the opacity of the \textit{r}-process elements, which does not produce polarization.
Of course, the electron scattering occurs in the \textit{r}-process element ejecta, but their opacity is much smaller than that of the nucleon ejecta \citep{Kasen+2013,Kyutoku+2015}.
In the \textit{r}-process element ejecta, electrons are supplied by the ionization of the \textit{r}-process elements, and the electron scattering opacity is given by
\begin{eqnarray}
\kappa_{\rm{es}}=\frac{x}{A}\kappa_{\rm{es,H}}\sim0.1\,\biggl(\frac{x}{10}\biggl)\biggl(\frac{A}{80}\biggl)^{-1}\,\kappa_{\rm{es,H}},
   \label{kappa_es for r}
\end{eqnarray}
where $x$, $A$, and $\kappa_{\rm{es,H}}=0.4\,\rm{cm^2\,g^{-1}}$ are the degree of ionization, the averaged atomic number of the \textit{r}-process elements, and Thomson scattering opacity of hydrogen, respectively.
Therefore, the fast-tail emission may show a much larger degree of polarization than that of the blue and red macronovae.
We also remark that the observed luminosity and timescale of the blue macronova in GW170817 suggest the ejecta mass and opacity of $M_{\rm blue}\simeq0.02\,\Msun$ and $\kappa_{\rm r}\simeq0.1-1{\,\rm cm^{2}\,g^{-1}}$ \citep[e.g.][]{Drout+2017}.
The rather small opacity supports our assumption that the electron scattering is still the main opacity source even if a fraction of \textit{r}-process elements are synthesized in the fast tail and mixed with free nucleons.

\begin{figure}
\begin{center}
\includegraphics[width=60mm, angle=270]{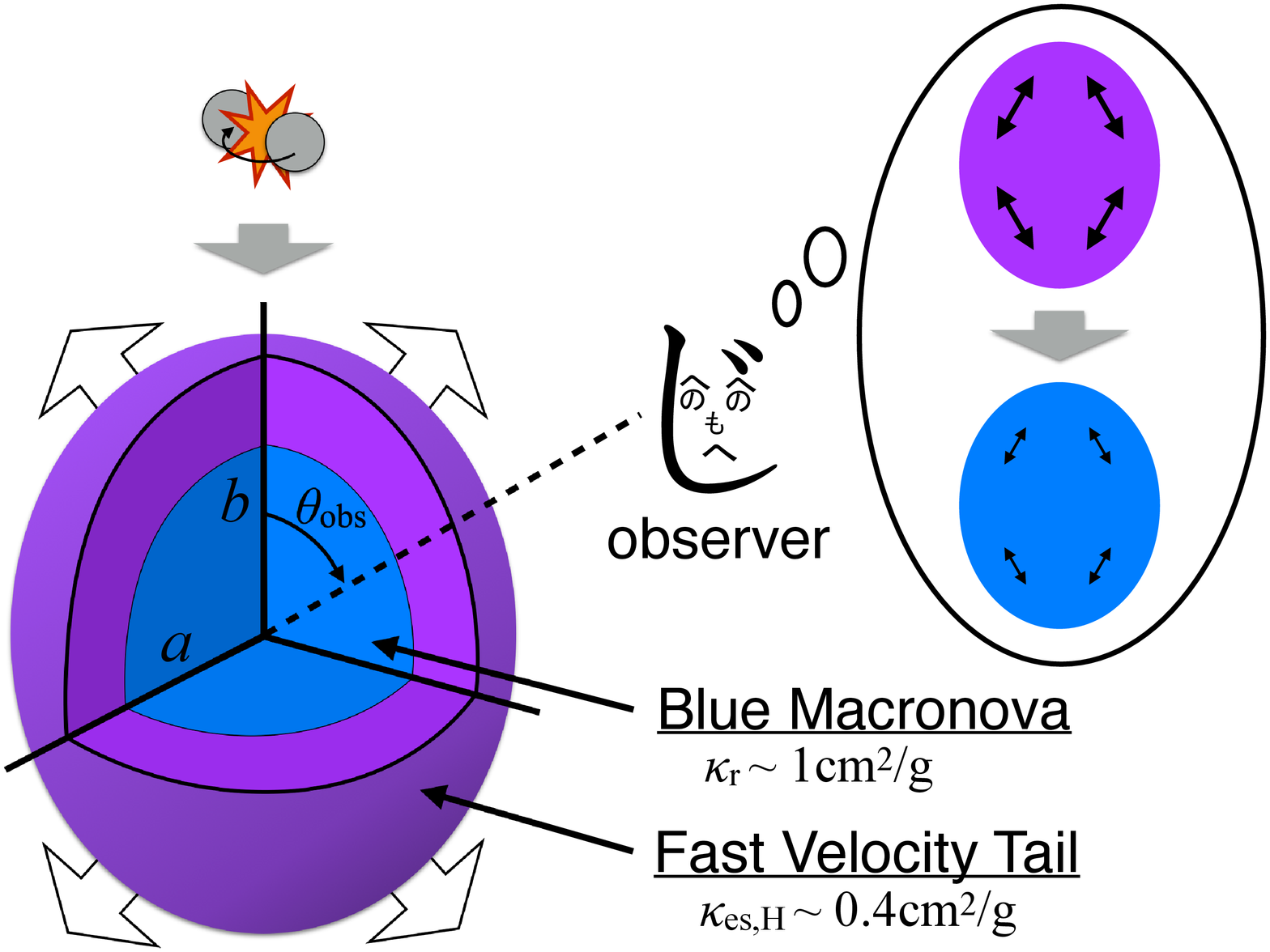}
\caption{A schematic picture of the configuration of the ejecta and their polarization.
A fast velocity tail (purple) is located at the outermost layer and expands faster than blue macronova ejecta (blue).
In the tail, nucleons remain and the electron scattering dominates the opacity, while the bound-bound transition of the \textit{r}-process elements is the main opacity source in the blue macronova ejecta.
The photosphere is assumed to be an ellipsoidal shape.
An observer sees the merger event with a viewing angle $\theta_{\rm{obs}}$.}
\label{fig picture}
\end{center}
\end{figure}

We can also expect the time evolution of the polarization degree.
When the photosphere is located in the fast velocity tail, where the electron scattering dominates the opacity, the degree of polarization can be large if the photosphere has a large asphericity.
As the photosphere recedes into the \textit{r}-process element-rich ejecta (e.g., blue or red macronova ejecta), the bound-bound transition dominates the opacity at the photosphere, which reduces the degree of polarization.
In Fig. \ref{fig picture}, we show a schematic picture of the situation we consider.

\subsection{Polarization}\label{polarization}
In the following, we demonstrate that a large polarization degree and temporal behavior can be produced in the first-hour macronovae.
The degree of polarization is defined by the ratio of the polarized intensity to the total intensity \citep{Chandrasekhar1960,Rybicki&Lightman1979}.
We approximately estimate the observed polarization degree $\Pi$ as
\begin{eqnarray}
\Pi\simeq\Pi_0\tau_{\rm{es}},
   \label{degree of polarization}
\end{eqnarray}
where $\Pi_0$ is the ``intrinsic'' degree of polarization which is made by a purely electron-scattering-dominated photosphere and gives the ratio of the intensities produced by the electron scattering at the photosphere, and $\tau_{\rm{es}}$ is the optical depth of the electron scattering at the photosphere which approximately gives the ratio of the photon intensity scattered by electrons and total intensity at the photosphere.
In order to calculate these quantities, we make the following assumptions.
(a) In the fast velocity tail, the \textit{r}-process elements are not synthesized, and free nucleons such as neutrons and protons remain.
When the photosphere is located in this nucleon layer, the electron scattering dominates the opacity.
(b) The high velocity tail has an aspherical morphology, which determines the intrinsic degree of polarization $\Pi_0$. 
We assume that the photosphere keeps its morphology and the intrinsic polarization degree is fixed.
In the latter of this subsection, we explain a method to evaluate the intrinsic degree $\Pi_0$.

The intrinsic degree of polarization $\Pi_0$ is the polarization produced by a purely electron-scattering-dominated photosphere.
An asphericity may be produced, for instance, when the fast velocity tail is ejected not in a spherically symmetric manner.
\cite{Hotokezaka+2018b} analyzed the results of the high-resolution numerical simulation by \cite{Kiuchi+2017}, and we can see that high velocity ejecta are produced anisotropically depending on the equation of state of the dense matter.
For simplicity, we approximate the photosphere as an ellipsoid.

The degree of polarization from oblate and prolate ellipsoids has been calculated in the context of SN explosions \citep{Shapiro&Sutherland1982,Hoflich1991,Wang&Wheeler2008}.
In this work, we evaluate the polarization degree of ellipsoids by using a result of Monte Carlo simulations by \cite{Hoflich1991}.
They calculated the degrees of polarization of ellipsoids for various configurations.
The results are well described by (see their Figs. 4 and 7)
\begin{eqnarray}
\Pi_0\propto\begin{cases}\bigl(1-\frac{a}{b}\bigl)\sin^2\theta_{\rm{obs}}&\text(a<b),\\
\bigl(1-\frac{b}{a}\bigl)\sin^2\theta_{\rm{obs}}&\text(b<a),
\end{cases}
   \label{intrinsic polarization}
\end{eqnarray}
where $\theta_{\rm{obs}}$, $a$, and $b$ are the viewing angle and the axis lengths of the ellipsoid (see Fig. \ref{fig picture}).
For a small asphericity $a/b\simeq1$ (sphere) or a face-on observer $\theta_{\rm{obs}}\simeq0\,\rm{deg}$, the degree vanishes due to the symmetry.
The viewing angle dependence $\Pi_0\propto\sin^2\theta_{\rm{obs}}$ is consistent with an analytical calculation by \cite{Brown&McLean1977}.
By using the result of \cite{Hoflich1991} (their Fig. 7), we normalize Eq. \eqref{intrinsic polarization} and estimate the degrees of polarization for various axis ratios and viewing angles.

\begin{figure}
\begin{center}
\includegraphics[width=60mm, angle=270]{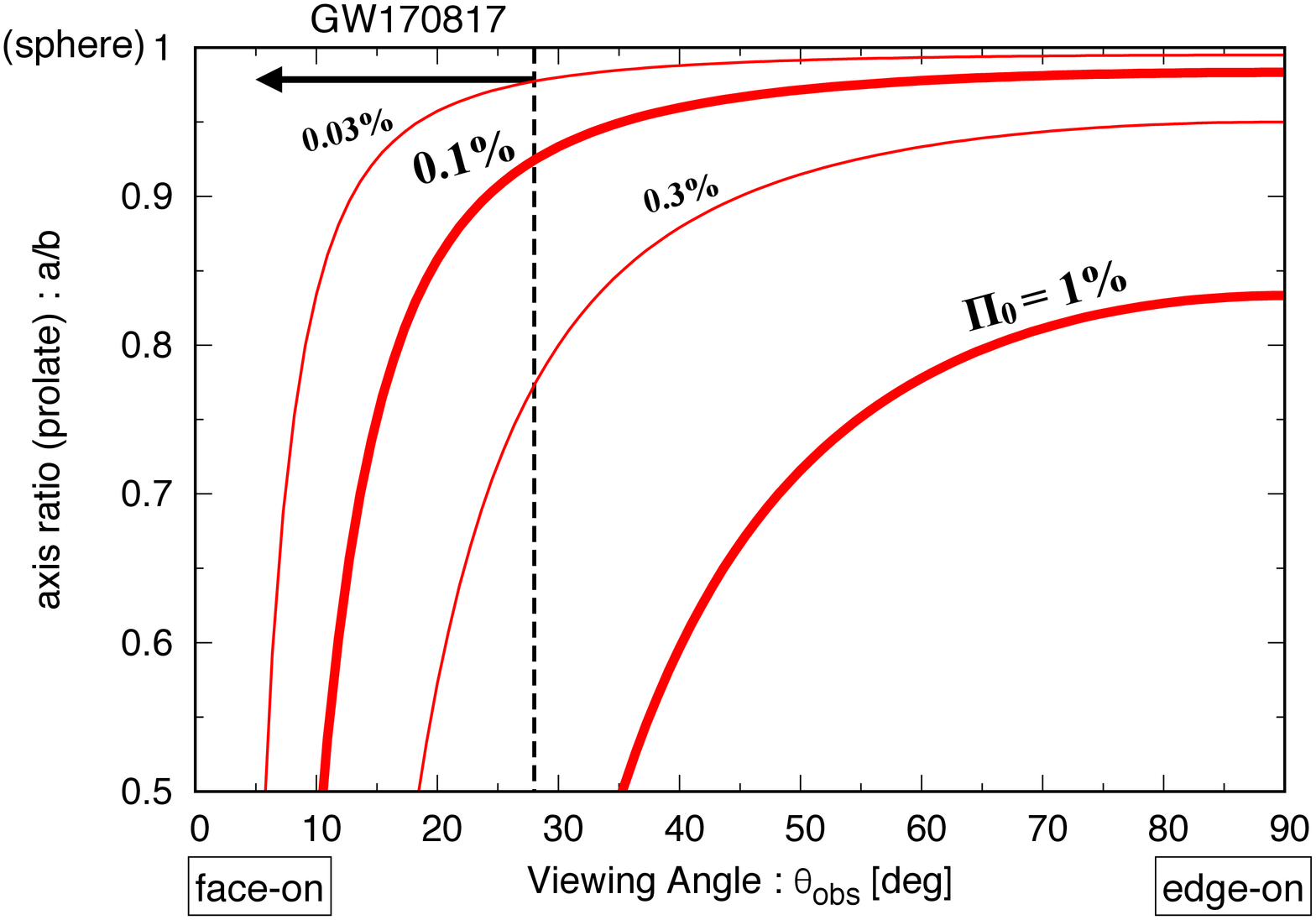}
\caption{Contours of the viewing angle and the axis ratio of ellipsoids which give the same degree of polarization for the purely electron-scattering-dominated photosphere.
The ellipsoid is prolate ($b<a$, see Fig. \ref{fig picture}).
The degrees of polarization vanishes for the sphere $a/b\simeq1$ or a face-on observer $\theta_{\rm{obs}}\simeq0{\,\rm{deg}}$ due to the rotational symmetry.
We also show the upper limit of the viewing angle of GW170817 with a dashed line.}
\label{fig pol pro}
\end{center}
\end{figure}

\begin{figure}
\begin{center}
\includegraphics[width=60mm, angle=270]{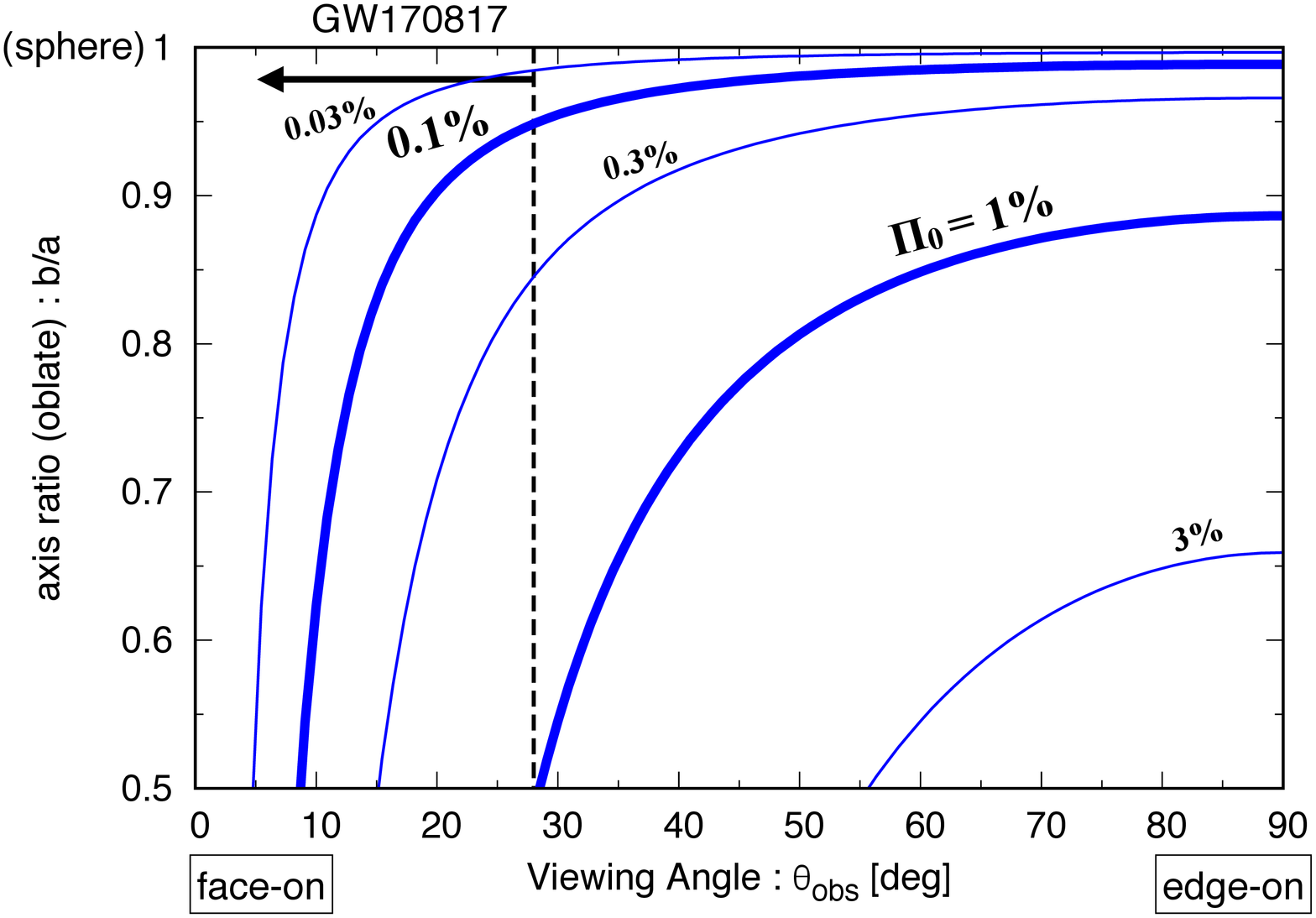}
\caption{The same as Fig. \ref{fig pol pro}, but for oblate ellipsoids.}
\label{fig pol ob}
\end{center}
\end{figure}

In Figs. \ref{fig pol pro} and \ref{fig pol ob}, we show the contours of the viewing angle and the axis ratio which give the same degree of polarization for the prolate ($a<b$) and oblate ($a>b$) ellipsoids, respectively.
The evaluated polarization degrees show similar values for prolate and oblate ellipsoids while the oblate shape shows a little larger polarization.
The maximum value of the polarization is about $\sim1-3\,\%$ for $a/b$ or $b/a\lesssim0.8$ and $\theta_{\rm{obs}}\gtrsim40\,\rm{deg}$.
With black dashed lines, we show the upper limit of the viewing angle of GW170817 $\theta_{\rm{obs}}\lesssim28\,\rm{deg}$ \citep{Abbott+2017c}, which is determined only by the GW observation.

\subsection{Optical depth and light curve}\label{light curve}
We calculate the electron-scattering optical depth at a photosphere and the light curve of the early macronovae, by assuming that ejecta are piecewise spherical.
This simplification may also be justified for our order-of-magnitude estimation.

The early macronovae can be powered by various mechanisms.
If the \textit{r}-process elements are synthesized in the head of fast velocity tail, their radioactive decay heating powers an early emission as late-time macronovae.
A free neutron precursor and a cocoon emission may also contribute to the early emission.
In this work, since we are interested in the polarization produced in the nucleon layer of the fast tail, we focus on the early macronovae powered by the \textit{r}-process element and free neutron radioactive decay heating.
It should be noted that the following results qualitatively hold for the cocoon powered case, which we discuss in section \ref{summary and discussion}.

We consider the fast tail and blue macronova ejecta components separately, and assume that they have power-law density profiles.
For the blue macronova component, numerical simulations have studied the density profile and found that it is reasonably approximated by a power-law profile \citep{Hotokezaka+2013,Nagakura+2014}.
For the fast velocity tail, the power-law density profile may be a reasonable approximation \citep{Kyutoku+2014,Hotokezaka+2018b}.

\begin{figure}
\begin{center}
\includegraphics[width=60mm, angle=270]{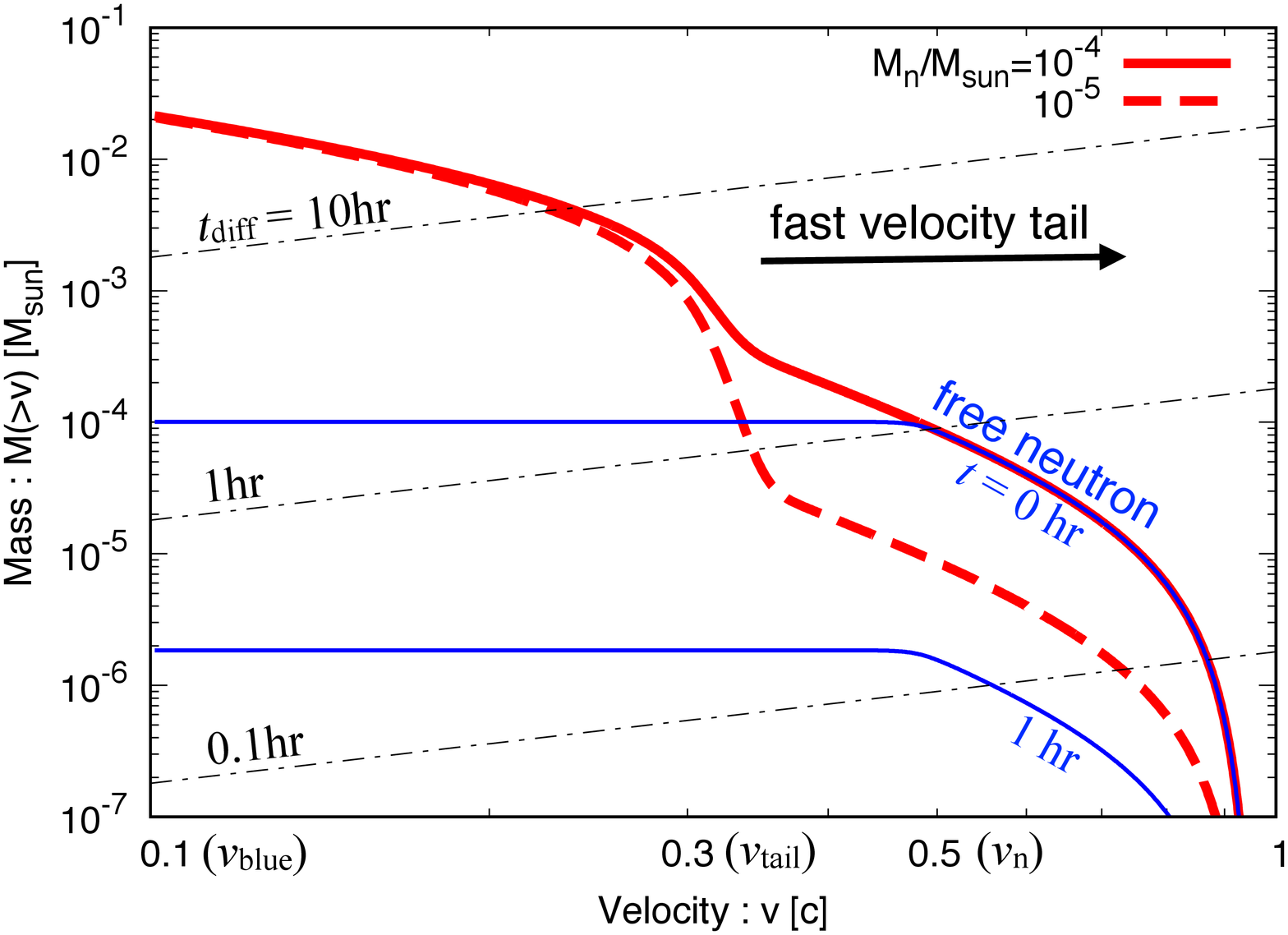}
\caption{Enclosed-mass profiles of the blue macronova ejecta ($v\lesssim{v_{\rm{tail}}}=0.3\,c$) and fast velocity tail ($v\gtrsim{}v_{\rm{tail}}$).
Red solid and dashed curves denote the profiles with the nucleon mass of $M_{\rm{n}}=10^{-4}$ and $10^{-5}\,\Msun$, respectively.
The blue solid curves show the enclosed mass of free neutrons at 0 and 1 hr after the merger where we assume that only neutrons compose the tail above $v>v_{\rm{n}}=0.5\,c$.
The black dash-dotted lines show the diffusion mass as a function of velocity at $t=0.1$, $1$, and $10\,\rm{hrs}$ from bottom to top.}
\label{fig profile}
\end{center}
\end{figure}

We locate the blue macronova ejecta and the fast velocity tail at the reasonable ranges of velocity and give their density profiles as
\begin{eqnarray}
\rho_{\rm{blue}}(v,t)&=&\frac{M_{\rm{blue}}}{4\pi(v_{\rm{blue}}t)^3}\biggl(\frac{v}{v_{\rm{blue}}}\biggl)^{-\beta_{\rm{blue}}}\biggl[\frac{\beta_{\rm{blue}}-3}{1-(\frac{v_{\rm{tail}}}{v_{\rm{blue}}})^{3-\beta_{\rm{blue}}}}\biggl]\nonumber\\
&&(v_{\rm{blue}}<v<v_{\rm{tail}}),\\
\rho_{\rm{tail}}(v,t)&=&\frac{M_{\rm{n}}}{4\pi(v_{\rm{n}}t)^3}\biggl(\frac{v}{v_{\rm{n}}}\biggl)^{-\beta_{\rm{tail}}}\biggl[\frac{\beta_{\rm{tail}}-3}{1-(\frac{c}{v_{\rm{n}}})^{3-\beta_{\rm{tail}}}}\biggl]\nonumber\\
&&(v_{\rm{tail}}<v),
\end{eqnarray}
respectively.
The quantities $M_{\rm{blue}}$, $v_{\rm{blue}}$, $\beta_{\rm{blue}}$, $M_{\rm{n}}$, $v_{\rm{tail}}$, $v_{\rm{n}}$, and $\beta_{\rm{tail}}$ are the total mass, the lowest velocity, and the power-law index of the blue macronova ejecta, the total nucleon mass, the lowest velocities of the tail and nucleon mass shell, and the power-law index of the fast tail's profile, respectively.
For the blue macronova ejecta, we set $M_{\rm{blue}}=2\times10^{-2}\,\Msun$, $v_{\rm{blue}}=0.1\,c$, $v_{\rm{tail}}=0.3\,c$, and $\beta_{\rm{blue}}=4$, for the fiducial values.
While the ejecta mass and highest velocity ($v_{\rm tail}$) of the blue macronova ejecta are motivated by the observations \citep[e.g.,][]{Cowperthwaite+2017,Drout+2017,Kilpatrick+2017}, the lowest velocity ($v_{\rm blue}$) of the ejecta is uncertain. However, the shell at the lowest velocity contributes the emission at later time $t\gtrsim{\rm\,days}$, which does not have an effect on the polarization signature at the early timescale.
The tail profile is normalized at the velocity $v_{\rm{n}}$ above which nucleons remain with a total mass $M_{\rm{n}}$ (see below, for the composition of the tail).
At each edge of the both profiles, we suppress the profiles with Gaussian functions.
In Fig. \ref{fig profile}, we show the distribution of the enclosed mass which is defined by
\begin{eqnarray}
M(>v)=\int_{vt}^{ct}dr\,4\pi{r^2}(\rho_{\rm{blue}}+\rho_{\rm{tail}}).
   \label{enclosed mass}
\end{eqnarray}
The red solid and dashed curves show the mass profile with $M_{\rm{n}}=10^{-4}$ and $10^{-5}\,\Msun$, respectively.
For the fast velocity tail, the parameters are adopted as $v_{\rm{n}}=0.5\,c$, which is motivated by \cite{Metzger+2015}, and $\beta_{\rm{tail}}=6$.
Different values of the index do not change our result so much as long as the enclosed mass is fixed at $v_{\rm{n}}$.
The distribution of the kinetic energy is also consistent with the numerical results by \cite{Hotokezaka+2018b} and the estimated value from the afterglows of GW170817 \citep{Matsumoto+2018c}.

We also make an assumption on the chemical composition in the fast tail in order to calculate the electron-scattering optical depth and the light curve of neutron precursors.
To see two extreme cases, we consider the following models along \cite{Metzger+2015}.
In one model, we assume that the tail is composed of only free neutrons as survived nucleons above $v>v_{\rm{n}}$.
We use the following mass fraction of the \textit{r}-process elements, free neutrons, and protons which are the products of the neutrons' beta-decay as
\begin{eqnarray}
X_{\rm{r}}(v,t)&=&\frac{2}{\pi}\arctan{\biggl[\biggl(\frac{M(>v)}{M_{\rm{n}}}\biggl)^\alpha}\biggl],
   \label{mass fraction r}\\
X_{\rm{n}}(v,t)&=&(1-X_{\rm{r}})e^{-t/t_{\rm{n}}},\\
X_{\rm{p}}(v,t)&=&(1-X_{\rm{r}})(1-e^{-t/t_{\rm{n}}}),
\end{eqnarray}
where $t_{\rm{n}}=900\,{\rm{s}}$ is the beta-decay timescale\footnote{There is a $3.9\,\sigma$ discrepancy between the neutron lifetime measured by using neutrons in a beam \citep[$t_{\rm{n}}=887.7\pm2.2{\,\rm{s}}$,][]{Yue+2013} and trapped neutrons \citep[$t_{\rm{n}}=878.5\pm0.8\,\rm{s}$,][]{Serebrov+2008}.
Unfortunately, this difference does not make a significant change in our results and we cannot use neutron precursors as an independent method to measure the neutron lifetime.} and the index $\alpha$ expresses how sharply the nucleon (neutrons, in this case) abundance decreases below $v<v_{\rm{n}}$.
While we set $\alpha=10$, other parameter value does not change the results so much.
In Fig. \ref{fig profile}, we also show the enclosed mass of the neutrons with blue curves at $t=0$ and $1\,\rm{hr}$ after the merger.
Due to the beta-decay, the amount of free neutrons decreases.
In the other model, we assume that nucleons do not survive and only the \textit{r}-process elements exist, $X_{\rm{r}}(v,t)=1$.
For the blue macronova ejecta, we also set $X_{\rm{r}}=1$.

For each mass shell, we can calculate the diffusion time given by Eq. \eqref{diffusion time}.
Conversely, we can estimate the position of the diffusion shell which powers the emission at time $t=t_{\rm{diff}}$.
In Fig. \ref{fig profile}, we also show the diffusion shell whose diffusion times are $t_{\rm{diff}}=0.1$, $1$, and $10\,\rm{hrs}$ with dash-dotted lines for $\kappa=\kappa_{\rm{es,H}}$.
Intersections of these lines with the enclosed-mass profile roughly give the mass shells which power the emission  at $t=t_{\rm{diff}}$.

We derive the diffusion shell and the photosphere in the ejecta for the above chemical composition.
These positions are given by solving the equations of $\tau=c/v$ and $\tau=1$, respectively, where $\tau$ is the optical depth.
As we discussed above, there are two opacity sources in the ejecta, e.g., the electron scattering and the bound-bound transition of the \textit{r}-process elements.
Then, the optical depth is given by
\begin{eqnarray}
\tau(v,t)=\int_{vt}^{ct}dr[\kappa_{\rm{r}}\rho_{\rm{blue}}+(\kappa_{\rm{r}}X_{\rm{r}}+\kappa_{\rm{es,H}}X_{\rm{p}})\rho_{\rm{tail}}],
   \label{optical depth}
\end{eqnarray}
where $\kappa_{\rm{r}}=1\,\rm{cm^2}\,g^{-1}$ is the bound-bound opacity of the \textit{r}-process elements in the blue macronova ejecta \citep{Tanaka+2018}.
In Eq. \eqref{optical depth}, we ignore the contribution from the electron scattering by electrons supplied by the ionization of the \textit{r}-process elements, which is justified due to its small opacity (Eq. \ref{kappa_es for r}).
We solve the equations $\tau=c/v$ and $\tau=1$ numerically and derive the velocity coordinates of the diffusion shell $v_{\rm{diff}}$ and photosphere $v_{\rm{ph}}$ at each time.

The bolometric luminosity is given by using the enclosed mass of the diffusion shell $M(>v_{\rm{diff}})$.
The mass of the \textit{r}-process elements and free neutrons in the enclosed mass are evaluated by
\begin{eqnarray}
M_{\rm{r}}(>v_{\rm{diff}})&=&\int_{v_{\rm{diff}}t}^{ct}dr\,4\pi{r^2}(\rho_{\rm{blue}}+X_{\rm{r}}\rho_{\rm{tail}}),\\
M_{\rm{n}}(>v_{\rm{diff}})&=&\int_{v_{\rm{diff}}t}^{ct}dr\,4\pi{r^2}X_{\rm{n}}\rho_{\rm{tail}}.
\end{eqnarray}
The bolometric luminosity is estimated by
\begin{eqnarray}
L\simeq{q_{\rm{r}}}M_r(>v_{\rm{diff}})+q_{\rm{n}}M_{\rm{n}}(>v_{\rm{diff}}),
\end{eqnarray}
where we use the specific heating rates of the radioactive decay of the \textit{r}-process elements and neutrons of \cite{Wanajo+2014,Kulkarni2005}
\begin{eqnarray}
q_{\rm{r}}&=&2\times10^{10}\biggl(\frac{t}{\rm{day}}\biggl)^{-1.3}\,\rm{erg\,s^{-1}\,g^{-1}},
   \label{q_r}\\
q_{\rm{n}}&=&3\times10^{14}\,\rm{erg\,s^{-1}\,g^{-1}},
\end{eqnarray}
respectively.
It should be noted that for blue macronova ejecta with $\kappa_{\rm r}\simeq0.1-1\,\rm cm^{2}\,g^{-1}$, which suggests thier initial electron fraction $Y_{\rm e}\gtrsim0.3$ \citep{Tanaka+2018}, Eq. \eqref{q_r} gives a correct heating rate up to $t\lesssim{\rm day}$, but overestimates the rate at later time \citep[see Fig. 5 in ][]{Wanajo+2014}.

At the photosphere $R_{\rm{ph}}=v_{\rm{ph}}t$, we evaluate the electron-scattering optical depth $\tau_{\rm{es}}$.
The electron scattering opacity at the photosphere is given by
\begin{eqnarray}
\tau_{\rm{es}}=\int_{v_{\rm{ph}}t}^{ct}dr\,\kappa_{\rm{es,H}}\biggl[\frac{x}{A}(\rho_{\rm{blue}}+X_{\rm{r}}\rho_{\rm{tail}})+X_{\rm{p}}\rho_{\rm{tail}}\biggl],
\end{eqnarray}
where we take the contributions from electrons supplied by the ionization of the \textit{r}-process elements into account (the first and second terms), which we ignored in Eq. \eqref{optical depth}.
For the mean mass number of the \textit{r}-process elements, we adopt $A=80$.
The degree of ionization of the \textit{r}-process elements $x$ is approximated by using the ionization degree of germanium (the most abundant \textit{r}-process element for the solar abundance) at the same density and the photospheric temperature in the photosphere, and given by solving the Saha equation. 
The photospheric temperature is given by $T_{\rm{ph}}=(L/4\pi\sigma_{\rm{SB}}R_{\rm{ph}}^2)^{1/4}$, where $\sigma_{\rm{SB}}$ is the Stefan-Boltzmann constant.

\section{results}\label{results}
\begin{figure}
\begin{center}
\includegraphics[width=60mm, angle=270]{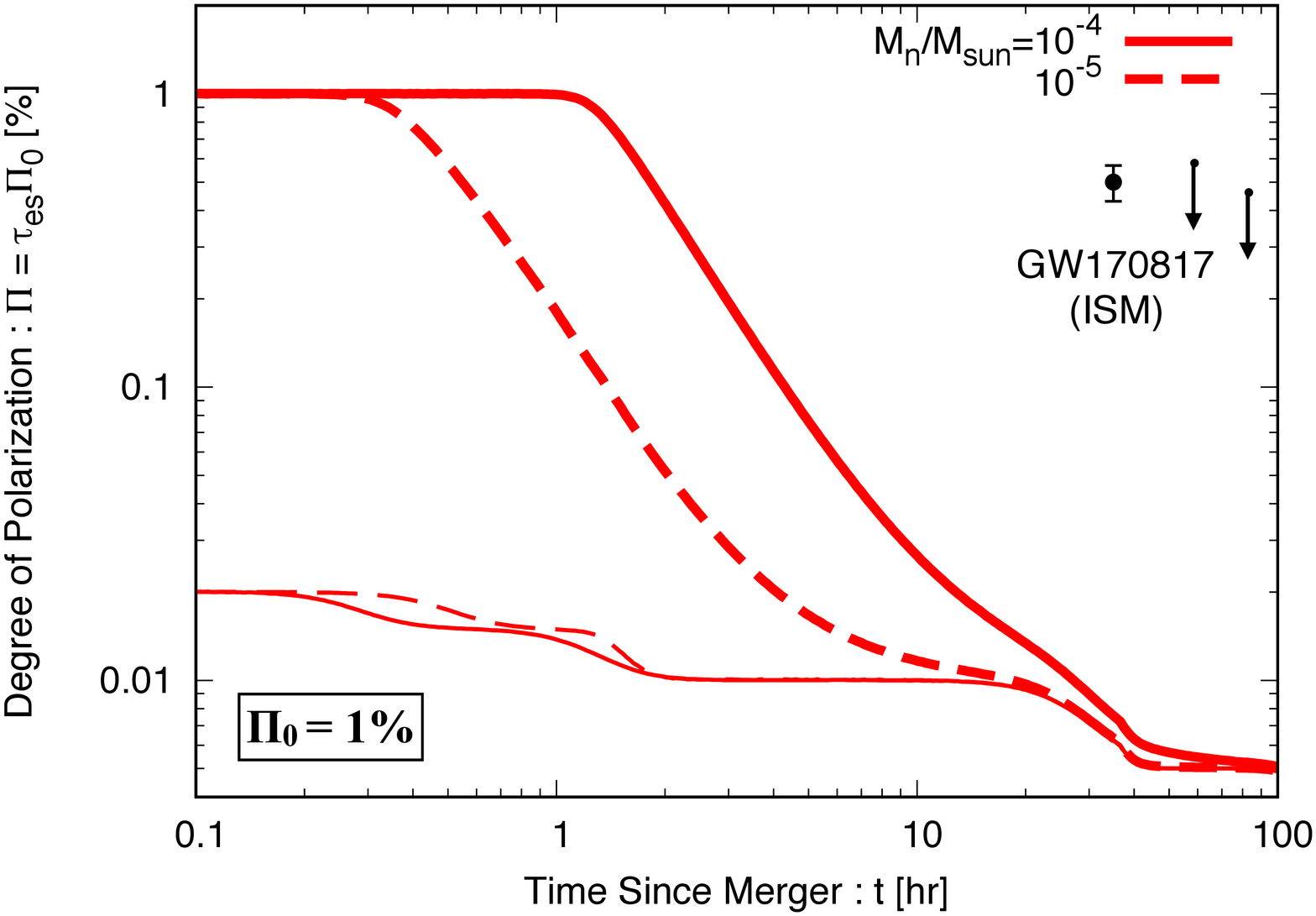}
\caption{Time evolution of the degree of polarization from the fast velocity tail and blue macronova ejecta.
The thick red solid and dashed curves show the polarization degrees of ejecta with $M_{\rm{n}}=10^{-4}$ and $10^{-5}\,\Msun$, respectively.
The thin curves denote the polarization degrees for ejecta with the same parameters as thick ones but no free nucleons ($X_{\rm{r}}=1$).
For these curves, we set the intrinsic polarization degree as $\Pi_0=1\,\%$.
The data points show the observed degree of polarization for GW170817 \citep{Covino+2017}, and the first point is consistent with the polarization produced by the dust scattering in the ISM.}
\label{fig tau}
\end{center}
\end{figure}

In Fig. \ref{fig tau}, we show the time evolution of the degree of polarization, which is given by Eq. \eqref{degree of polarization}.
We set the intrinsic polarization degree as $\Pi_0=1\,\%$, which results in that the polarizations shown in Fig. \ref{fig tau} are equal to the electron-scattering optical depths $\tau_{\rm{es}}$.
The thick red solid and dashed curves show the polarizations for the ejecta with free neutrons of $M_{\rm{n}}=10^{-4}$ and $10^{-5}\,\Msun$, respectively.
At first, both of the ejecta show the intrinsic polarization degree $\Pi=\Pi_0$, but the polarizations decay soon about $1\,\rm{hr}$ ($M_{\rm{n}}=10^{-4}\,\Msun$) and $0.3\,\rm{hr}$ ($M_{\rm{n}}=10^{-5}\,\Msun$) after the merger.
This temporal behavior is explained as follows.
At the early timescale, the photosphere is located in the free neutron layer and the electron-scattering optical depth is unity.
As the photosphere recedes to the \textit{r}-process elements-rich ejecta ($v<v_{\rm{n}}$), the electron scattering is suppressed, which results in the significant decrease of the polarization degree.
In this phase, the polarization degree decreases as $\Pi\propto{t^{-2}}$ tracing the electron-scattering optical depth $\tau_{\rm{es}}\simeq\kappa_{\rm{es,H}}M_{\rm{n}}/4{\pi}R^2\propto{t^{-2}}$.
Since the degree of polarization decays faster for the ejecta with smaller total neutron mass, the timescale may be useful to evaluate the total neutron mass.

The thin red solid and dashed curves show the time evolution of the polarization degrees for the ejecta with the same parameters as the thick ones but $X_{\rm{r}}=1$ (no free neutrons).
Due to the large mass number $A$, the electron scattering is not the main opacity source in these \textit{r}-process element-rich ejecta (see Eq. \ref{kappa_es for r}).
In this case, the observed polarization is much smaller than the intrinsic polarization degree.
The small drops at $t\simeq0.3$, $1$, and $20\,\rm{hrs}$ in these curves are caused by the recombination of electrons to the \textit{r}-process elements (the decrease of the ionization degree $x$).

We also show the polarimetric observation data of GW170817 with black points \citep{Covino+2017}.
The first observation was carried out at $1.46\,\rm{days}\simeq35\,hrs$ after the merger.
The observed value is consistent with the polarization produced by the dust scattering in the interstellar medium (ISM).
If the polarimetric observation is performed at the early timescale $t\lesssim{\,\rm{hr}}$, we may detect the intrinsic polarization degree.

\begin{figure}
\begin{center}
\includegraphics[width=60mm, angle=90]{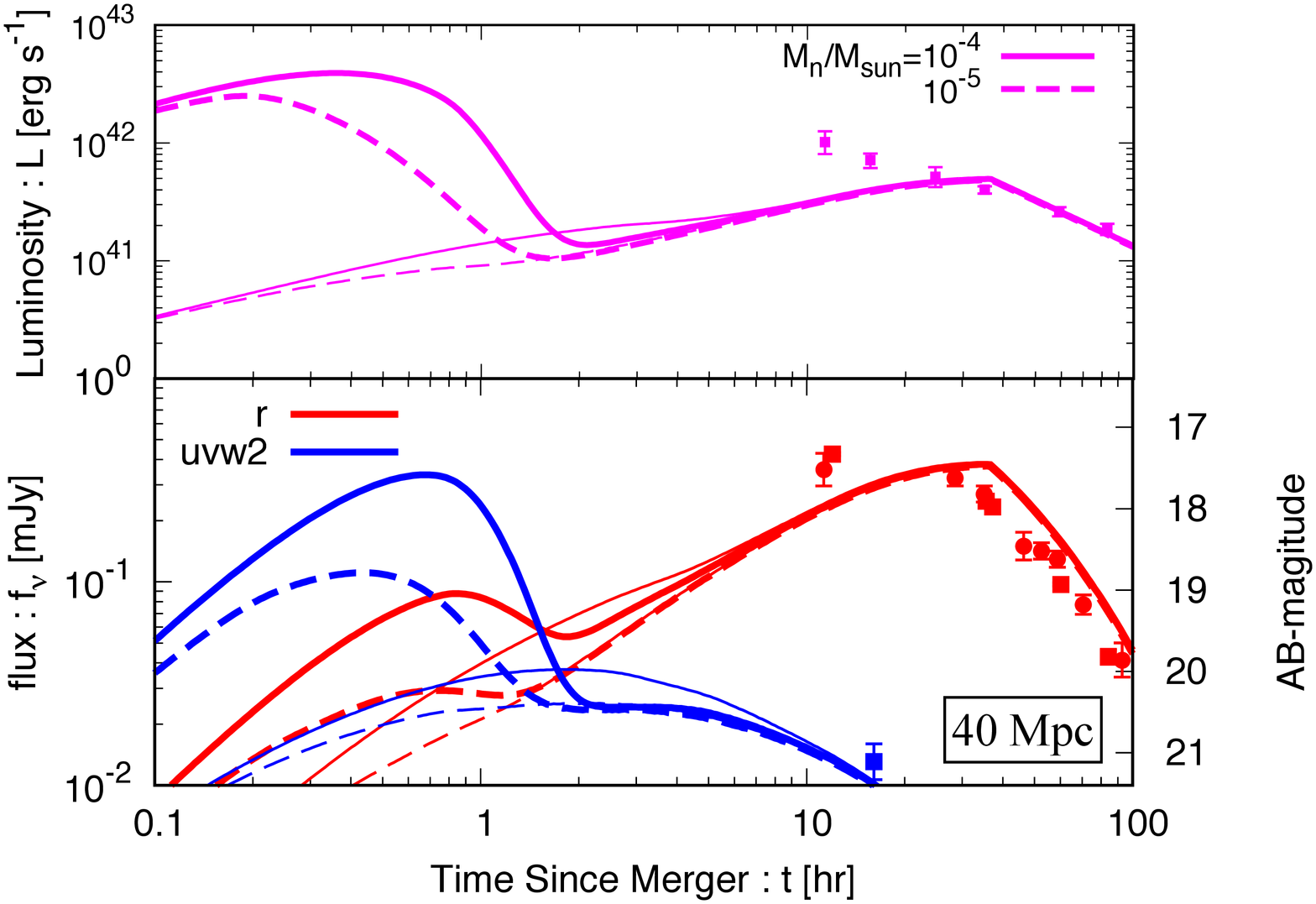}
\caption{The bolometric (top) and optical and UV bands' (bottom) light curves of the fast velocity tail ($\lesssim{\,\rm{hr}}$) and the blue macronova ejecta  ($\gtrsim{\,\rm{hr}}$).
The thick solid and dashed curves show the light curves for $M_{\rm{n}}=10^{-4}$ and $10^{-5}\,\Msun$, respectively.
In the bottom panel, red and blue curves show the r-band (optical) and uvw2-band (UV) light curves, respectively.
Compared with the no-neutron cases (thin curves), the free-neutron decay produces a bump at $t\lesssim\rm{\,hr}$ as a neutron precursor \citep[see][]{Metzger+2015}.
The points are the observation data of GW170817 taken from \citep{Kilpatrick+2017} (top), \citep{Kasliwal+2017} (bottom, circle), and \citep{Drout+2017} (bottom, square).}
\label{fig Lf}
\end{center}
\end{figure}

In Fig. \ref{fig Lf}, we show the light curves of the bolometric luminosity (top) and the r-band ($\nu=5\times10^{14}\,\rm{Hz}$) and uvw2-band ($\nu=1.5\times10^{15}\,\rm Hz$) fluxes (bottom) from the ejecta with magenta, red, and blue curves, respectively.
The thick solid and dashed curves show the light curves for the ejecta of $M_{\rm{n}}=10^{-4}$ and $10^{-5}\,\Msun$, respectively.
The thin curves denote the light curves when no free neutrons remain and the emissions are powered only by the \textit{r}-process elements.
In the top panel, the data points show the macronova in GW170817 taken from \cite{Kilpatrick+2017}.
At the early time ($t\lesssim\rm{\,hr}$), the beta-decay heating produces a bump in the bolometric light curves.
This behavior is basically consistent with \cite{Metzger+2015}.
It should be noted that if a cocoon breaks out of ejecta, its cooling emission may also produce a similar bump in the light curve.

In the bottom panel, the light curves in the r-band and uvw2-band are depicted by setting the distance to the merger events as $40\,\rm{Mpc}$.
The data points show the observed light curve of GW170817 taken from \cite{Kasliwal+2017} (circle) and \cite{Drout+2017} (square).
In the r-band, the bumps are not as large as in the bolometric light curve.
This is because the photosphere is small $R_{\rm{ph}}\lesssim5\times10^{13}\,\rm{cm}$ at the early time and the photospheric temperature is as high as $T_{\rm{ph}}\gtrsim3\times10^{5}{\,\rm{K}}$.
Therefore, the emission is produced mainly in UV bands.
Actually, the flux in uvw2-band is $\gtrsim3$ times larger than the r-band flux and the bumps are pronounced in UV bands.
The small flux in optical bands makes the optical polarimetric observations challenging (see section \ref{summary and discussion}).

\section{Summary \& Discussion}\label{summary and discussion}
In this work, we study the degree of polarization from early macronovae at the first hour, and its time evolution.
We find that if a fast expanding outermost layer of ejecta is aspherical and has a sufficient amount of free nucleons such as neutrons, the degree of polarization becomes larger than that of the ejecta composed of the \textit{r}-process elements only.
For the total nucleon mass of $10^{-5}$ and $10^{-4}\,\Msun$, the degree of polarization shows the ejecta's ``intrinsic'' value $\Pi_0$, which is dictated by the ejecta morphology, for the first $\sim0.3$ and $1\,\rm{hr}$, respectively.
We evaluate the intrinsic polarization degree by assuming the ejecta as an ellipsoid.
For a reasonable range of the axis ratio of the ellipsoid ($a/b$ or $b/a>0.5$), the intrinsic polarization has a value of $\Pi_0\lesssim3\,\%$.
The degree drops when the photosphere passes through the nucleon layer.
The timescale of the drop is given by the transparent timescale of the free nucleon shell as \citep{Kisaka+2015,Matsumoto+2018b}
\begin{eqnarray}
t_{\rm{tr}}=\sqrt{\frac{\kappa{M_{\rm{n}}}}{4\pi{v^2}}}.
   \label{photosphere timescale}
\end{eqnarray}
After the passage, the bound-bound transition of the \textit{r}-process elements dominates the opacity and reduces the degree of polarization as $\Pi\propto{t^{-2}}$. 

The detection of the polarization from early macronovae tells us not only the morphology of the ejecta but also the amount of free nucleons which survived the \textit{r}-process nucleosynthesis.
As shown in Figs. \ref{fig pol pro} and \ref{fig pol ob}, the intrinsic polarization is a function of the viewing angle and the axis ratio.
For binary NS merger events, we can constrain the viewing angle by the GW observation only \citep{Abbott+2017c}, or combining it with the electromagnetic observations \citep{Mandel2018,Finstad+2018,Mooley+2018b}.
A precise determination of the viewing angle helps us to constrain the morphology of the early macronovae.
The total free nucleon mass is evaluated by the time when the polarization decreases significantly (Eq. \ref{photosphere timescale}).
Since we can estimate the ejecta velocity by measuring the bolometric luminosity and photospheric temperature, we evaluate the total mass as
\begin{eqnarray}
M_{\rm{n}}\simeq\frac{4\pi{v^2t_{\rm{tr}}^2}}{\kappa}\simeq5\times10^{-5}\,\Msun\,\biggl(\frac{v}{0.5\,c}\biggl)^2\biggl(\frac{t_{\rm{tr}}}{1\,{\rm{hr}}}\biggl)^2\biggl(\frac{\kappa}{\kappa_{\rm{es,H}}}\biggl)^{-1}.
\end{eqnarray}

If a cocoon breaks out of the fast velocity tail, its cooling emission also contributes to the early macronovae \citep{Nakar&Piran2017,Gottlieb+2018,Wang&Huang2018}.
In this case, when nucleons remain at the fast tail and enrich the cocoon's outer layer, the detectable degree of polarization may be produced.
The cocoon light curve is dictated by the internal-energy (or temperature) distribution in the cocoon \citep{Nakar&Piran2017}, which depends on the complicated dynamics of the jet propagation and the cocoon.
Therefore, we need numerical calculations to follow the cocoon shape and the nucleon mixing at the surface, which is an interesting future work.
However, we can check whether an early emission is produced by a cocoon or not, by polarimetric observations.
When the early macronova shows a bump in the light curve, we can evaluate the required free neutron mass in the neutron precursor scenario.
On the other hand, we can also evaluate the nucleon mass from the timescale (Eq. \ref{photosphere timescale}) by polarimetric observations, which gives the upper limit on the total nucleon mass.
If the former evaluated mass is larger than the latter one, the early macronova requires the contribution from the cocoon. 

As shown in Fig. \ref{fig tau}, polarimetric observations should be performed within $\sim\,\rm{hr}$.
Such quick observations are actually carried out for some GRB afterglows \citep{Uehara+2012,Mundell+2013,Covino&Gotz2016}.
To carry out similar observations for binary mergers, the positions of the events have to be specified and alerted soon after the GW detections, which may be possible in the future observations.
However, it should be noted that the quick followups for GRB afterglows have been performed in optical bands.
As we show in the bottom panel of Fig. \ref{fig Lf}, due to the high photospheric temperature, the optical flux is not so large as the blue macronova emissions, which makes the accurate polarimetric observations hard.
For instance, for a source with flux $\sim0.1\,\rm{mJy}$ ($\sim19\,\rm{mag}$) in the r-band, an observation time of $\simeq10^3\,\rm{s}$ is required in order to detect the polarization degree with $1\,\%$ of accuracy by using the Kanata Telescope \citep{Kawabata+2008} or Liverpool Telescope \citep{Steele+2010}.
UV polarimetric facilities are required to observe the polarization degree from the early macronovae.

We discuss the uncertainties of our calculation.
First, we derived the intrinsic degree of polarization $\Pi_0$ only for the ellipsoidal photospheres and assumed that the polarization degree is fixed in time.
However, an actual ejecta-shape may have a more complicated structure and change temporally due to the multidimensional effect, an inhomogeneous chemical composition, and so on.
It is an interesting future work to calculate the polarization degree taking these effect into account.

Second, we assumed that the top of the tail ($v>v_{\rm{n}}$) is composed of only nucleons $X_{\rm{n}}+X_{\rm{p}}=1$ as an extreme case, which makes the electron-scattering optical depth $\tau_{\rm{es}}$ unity at this layer.
When the \textit{r}-process elements are synthesized and located in the outermost layer, it of course reduces the relative contribution of the electron scattering to the total optical depth at the photosphere.
For instance, the mass fraction of the \textit{r}-process elements is larger than $X_{\rm{r}}>0.5$ at $v>v_{\rm{n}}$, the optical depth of the electron scattering becomes $\tau_{\rm{es}}\lesssim0.3$ at the photosphere.
In this case, the polarization degree may be initially small and does not show a detectable reduction.

Finally, we calculated the light curve of a neutron precursor by setting the initial neutron-mass fraction as unity $X_{\rm{n}}=1$, which gives the maximum luminosity of the neutron precursor.
Realistically, the neutron mass fraction is not unity even in NSs before the merger and gets small due to the positron capture and helium synthesis \citep{Metzger+2015,Ishii+2018} and the precursor luminosity decreases.
The dimmer precursor may be helpful to perform polarimetric observations in optical bands owing to the smaller temperature.
It should be noted that if protons and helium rather than neutrons are produced, the electron scattering dominates the opacity in the outer layer.

\section*{Acknowledgements}
We thank Koutarou Kyutoku, Masaomi Tanaka, and Kenji Toma for reading the manuscript and giving us useful comments.
We are also grateful to Sho Fujibayashi and Nobuya Nishimura for fruitful discussions on the nucleosynthesis.
This work is supported by Grant-in-Aid for JSPS Research Fellow 17J09895 and JSPS Overseas Challenge Program for Young Researchers.

\bibliographystyle{mnras}
\bibliography{reference_matsumoto}

\bsp	
\label{lastpage}
\end{document}